# Two-photon interference from a bright single photon source at telecom wavelengths

*Je-Hyung Kim,*[†] *Tao Cai,*[†] *Christopher J. K. Richardson,*[‡] *Richard P. Leavitt,*[‡] *and Edo Waks*[*,†,§]

[†]Department of Electrical and Computer Engineering and Institute for Research in Electronics and Applied Physics, University of Maryland, College Park, Maryland 20742, USA

[‡]Laboratory for Physical Sciences, University of Maryland, College Park, Maryland 20740, USA

[§]Joint Quantum Institute, University of Maryland and the National Institute of Standards and Technology, College Park, Maryland 20742 , USA

Long-distance quantum communication relies on the ability to efficiently generate and prepare single photons at telecom wavelengths. In many applications these photons must also be indistinguishable such that they exhibit interference on a beamsplitter, which implements effective photon-photon interactions. However, deterministic generation of indistinguishable single photons with high brightness remains a challenging problem. We demonstrate two-photon interference at telecom wavelengths using an InAs/InP quantum dot in a nanophotonic cavity. The cavity enhances the quantum dot emission, resulting in a nearly Gaussian transverse mode profile with high out-coupling efficiency exceeding 36% after multi-photon correction. We also observe Purcell enhanced spontaneous emission rate up to 4. Using this source, we generate linearly polarized, high purity single photons at 1.3 μm wavelength and demonstrate the indistinguishable nature of the emission using a two-photon interference measurement, which exhibits indistinguishable visibilities of 18% without post-selection and 67% with post-selection. Our results provide a promising approach to generate bright, deterministic single photons at telecom wavelength for applications in quantum networking and quantum communication.

---

[*] Email: edowaks@umd.edu

Single photon sources are important building blocks for optical quantum information processing [1-4]. They are essential to generate photonic quantum bits (qubits) that can travel long distances over optical fibers and interconnect distant quantum network nodes [5-7]. Efficient on-demand single photon sources also enable quantum computation schemes based on either linear [3, 4] or nonlinear [8] optical elements.

Many applications in quantum communication require deterministic single-photon sources that emit at telecom wavelengths. Parametric down-conversion sources can operate in this wavelength range [9, 10] but provide only heralded single-photon states and cannot be easily extended to on-demand operation. In contrast, single quantum emitters provide the potential for creating on-demand single-photon sources [11, 12]. Quantum dots in III-V semiconductors are particularly promising quantum emitters that generate single photons with high indistinguishability at near-infrared wavelengths [13-18], and are also compatible with electrical injection [19, 20] and integration with nanophotonic structures [21-24]. A number of works have extended the emission of III-V quantum dots to telecom wavelengths by optimizing materials and growth parameters [25-31]. However, an on-demand source of indistinguishable single photons remains an outstanding challenge at telecom wavelength.

In this work, we demonstrate two-photon interference from a bright single photon source at telecom wavelengths. We use a single InAs/InP quantum dot in a photonic crystal cavity to attain bright and highly polarized single-photon emission at telecom wavelengths. Rather than using the fundamental mode of the cavity, we utilize a higher-order mode that exhibits better directionality and emits a transverse mode that can efficiently couple to fibers. Cavity-coupled single quantum dots exhibit significantly enhanced brightness with their high out-coupling efficiency. We also observe a Purcell enhancement of up to 4 for the spontaneous emission rate. From these cavity-coupled dots, we demonstrate the indistinguishability by performing Hong-Ou-Mandel-type interference measurements [32], which exhibits a clear two-photon interference effect. These results show that InAs/InP quantum dots coupled to photonic crystals can serve as bright indistinguishable single-photon sources for applications in long-distance quantum communication.

Figure 1(a) shows a scanning electron microscope image of a typical photonic crystal cavity structure used in this work. The cavity design is based on an L3 defect cavity [33] with a lattice parameter (denoted as a) of 370 nm, a hole radius of 0.27 a, and a slab thickness of 280 nm. To optimize the *Q* value of the fundamental cavity mode, we shift the three holes adjacent to the cavity outward by 0.26 a, 0.16 a, and 0.16 a, respectively.

To fabricate the device, we began with a wafer consisting of a 280 nm thick InP layer grown on top of a 2 μm-thick AlInAs sacrificial layer using molecular beam epitaxy. The center of the InP layer contained a thin layer of InAs quantum dots with densities of approximately 10 $\mu m^{-2}$. We have previously reported the details of the quantum dot growth procedure and material characterization [34]. We used plasma-enhanced chemical vapor deposition to deposit a 100-nm-thick silicon nitride thin film that served as an etching mask. We then patterned the mask using electron beam lithography and fluorine-based reactive ion etching, and transferred the pattern from the etch mask to the InP membrane using chlorine-based reactive ion etching. Finally, we removed the sacrificial layer by selective wet etching to form an air-suspended photonic crystal membrane.

We used a confocal microscope system as shown in Fig. 1(b) to optically characterize the device. We cooled the sample to 4 K using a low vibration closed cycle cryostat and excited the sample with a 780 nm laser emitting in either continuous wave or pulsed modes. We performed both excitation and collection using confocal microscopy with an objective lens that has a numerical aperture of 0.7. We used a spectrometer with a liquid nitrogen cooled InGaAs array to measure the spectrum of the emission and to select the desired quantum dot line. We also used a half-waveplate and a polarizing beamsplitter for polarization analysis of the collected emission. To perform time-resolved lifetime and photon correlation measurements, we used a 780 nm pulsed laser excitation source with a pulse width of 50 ps and a repetition rate of 40 MHz. For photon correlation measurements, we detected the photons using two fiber-coupled InGaAs single photon avalanche diodes.

We performed indistinguishability measurements of emitted photons using a Hong-Ou-Mandel-type two-photon interferometer. The interferometer was composed of two unbalanced fiber Mach-Zehnder interferometers. We coupled the pump laser to the first interferometer, composed of two fibers with a one meter path length difference to generate double pulses separated by 5 ns that excite the quantum dot. We then coupled the collected emission to the second interferometer which had the same path length difference to generate a two-photon interference effect.

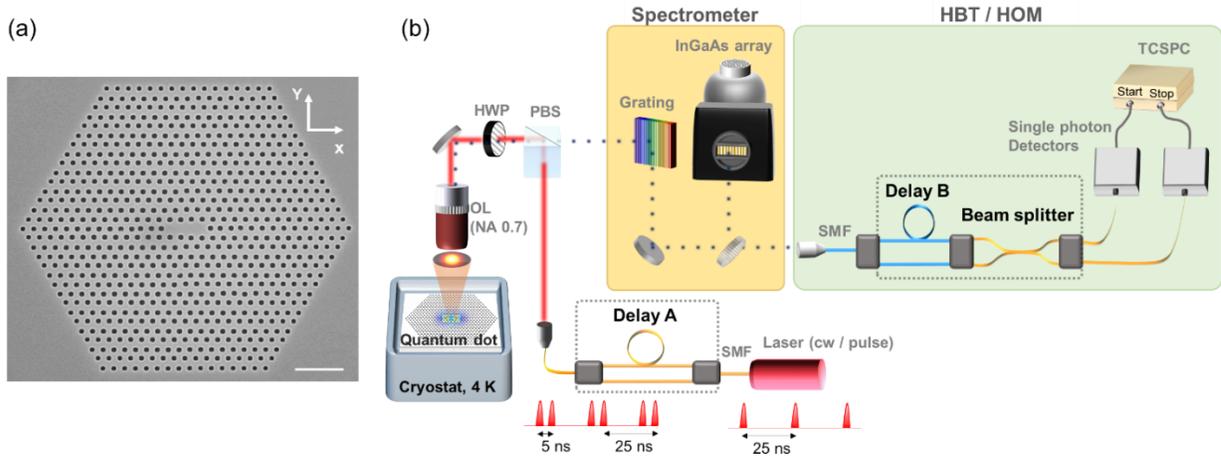

Fig. 1. (a) Scanning electron microscopy image of an air-suspended L3 photonic crystal cavity (scale bar is 2 μm) and (b) Schematic of measurement set-up. For single photon measurement by using a fiber-coupled Hanbury-Brown and Twiss (HBT) setup, the parts of delay A and B and beam splitter (dotted boxes in Fig. 1(b)) are removed from the setup. For two photon interference measurement by using a fiber-based Hong-Ou-Mandel (HOM) setup, two unbalanced Mach-Zehnder interferometers (dotted boxes) are inserted in both excitation and emission paths. OL, HWP, PBS, TCSPC, and SMF represent objective lens, half-wave plate, polarizing beam splitter, time-correlated single photon counter, and single mode fiber, respectively. Blue-colored SMF is a polarization maintained SMF.

We characterized the mode structure of the fabricated device using photoluminescence measurements. We excited the sample with a continuous wave laser at a high pump power of 30 µW in order to saturate all the quantum dots and measured the cavity emission spectrum, shown in Fig. 2(a). The spectrum exhibits several peaks corresponding to different cavity modes. The fundamental mode (denoted M1) occurs at a wavelength of 1380 nm and has a $Q$ of 7,000. In addition to the mode, we observe several higher order modes (M2-M5) that have $Q$ values of 4300, 380, 1000, and 2000, respectively. Figure 2(b) displays the magnitude of the simulated electric fields of the different cavity modes. Modes M1, M2, and M5 are polarized in the y direction (as indicated in Fig. 1(a)), while modes M3 and M4 are polarized in the x direction.

The high $Q$ value of mode M1 enables strong interactions and a large Purcell effect [35, 36]. However the collected emission from this mode is very weak due to the poor directionality of it's far-field pattern as shown in Fig. 2(c) [37]. The majority of the radiation emits at large angles that are outside the collection aperture of the objective lens (NA=0.7), denoted by the white circle, which leads to low collection efficiency. Furthermore, even when using a collection lens with a larger numerical aperture, the collected mode would have a transverse mode profile that is very difficult to couple to a single-mode fiber.

The higher-order modes of the photonic crystal cavity have far-field transverse mode profiles that overlap much better with the collection area of the lens [38, 39]. Mode M3 in particular has a Gaussian-like transverse mode pattern that is well-suited for coupling to a single-mode fiber. From the simulations we calculate that more than 80% of photons emitted from the top of the device lie within the acceptance angle of our collection lens, as compared to only 30 % for mode M1. We note that these numbers only consider the light radiated in the upward direction, and that additional losses occur due to emission of a fraction of the light below the sample.

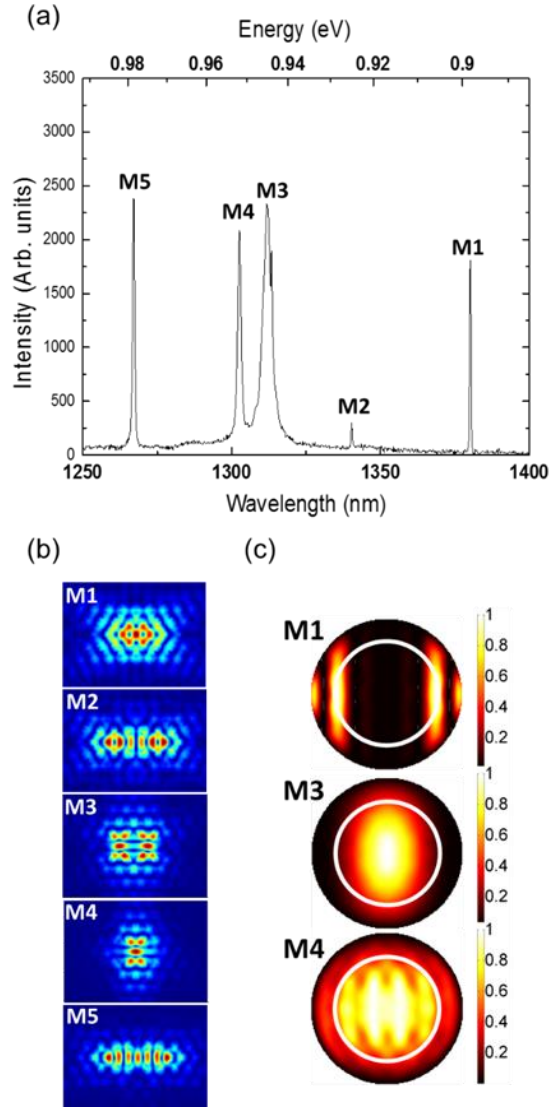

Fig. 2. Cavity mode analysis of L3 photonic crystal. (a) Cavity emission spectrum showing several cavity modes labeled M1-M5. (b) Simulated electric field profiles ($|E|$) of the cavity modes M1-M5 shown in (a). (c) Simulated far-field patterns of M1, M3, and M4 modes. x-dipole (y-dipole) sources and $|E_x|^2$ ($|E_y|^2$) components were used for far-field simulation of M3 and M4 modes (M1 mode). The white circle represents the collection angle θ = 45°, corresponding to the objective lens with NA = 0.7.

To observe single quantum dot emission we now reduced the laser excitation power to 100 nW in order to avoid saturation and power broadening. Figure 3(a) shows the photoluminescence spectrum at this pump power from a cavity region (black line) as well as a region away from the cavity (red line). The emission from the cavity region exhibits a clearly enhanced brightness as compared to the bulk. Near the resonance of mode M3, we observe a bright narrow line corresponding to a single quantum dot emission which we label as dot A. This quantum dot line has a spectrometer-resolution-limited linewidth of 50 μeV at 1320 nm, which matches the fiber optical *O*-band.

To confirm the single-photon nature of the cavity-coupled dot A, we performed second-order correlation measurements using a Hanbury-Brown and Twiss intensity interferometer. We filtered the quantum dot emission using a spectrometer, and detected the filtered emission using a 50:50 fiber beamsplitter and two InGaAs single photon avalanche diodes. We excited the quantum dot by using a 780 nm pulsed laser with a repetition rate of 40 MHz, a pulse width of 50 ps., and an average power of 5 nW.

Figure 3(b) shows the measured second-order correlation ($g^{(2)}(\tau)$), which exhibits a clear antibunching at $\tau = 0$. We fit the correlation curve to two-sided exponential functions convolved with a Gaussian function that accounts for the limited detector response time of 200 ps. We measured the detector dark counts and used this value as a background level (gray-colored region) of the fitted curve (See Supplement 1). From the fit we determine $g^{(2)}(0)=0.085\pm0.022$, indicating the bright emission of dot A originates from a single quantum dot with highly suppressed two-photon emission. The value of $g^{(2)}(0)$ increases with excitation power, but it remains below 0.5 up to the maximum quantum dot intensity (See Supplement 3).

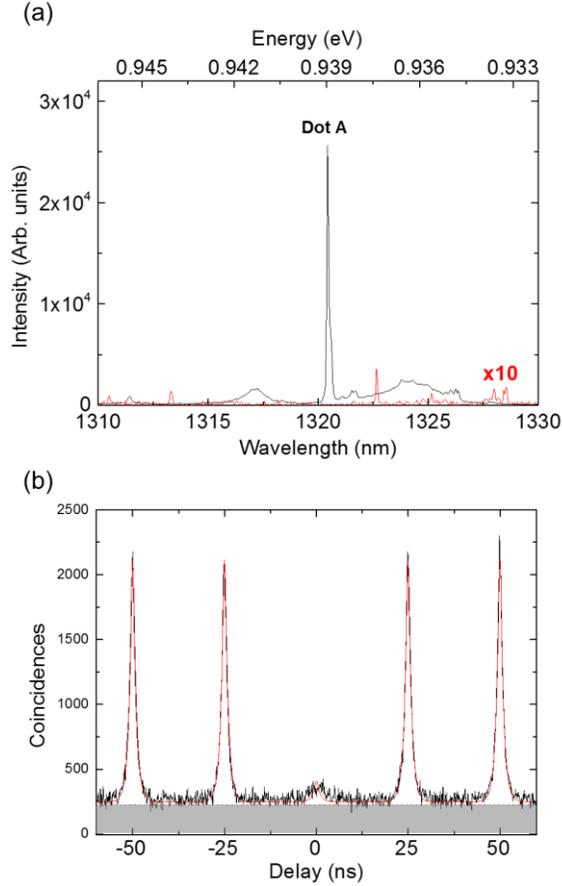

Fig. 3. Single photon emission from the cavity-coupled single quantum dot. (a) Photoluminescence spectra of the quantum dots coupled to mode M3 (black line) and the bulk quantum dots outside of a photonic crystal structure (red line). For comparison, the intensity of the bulk quantum dots is multiplied by a factor of 10. (b) Second-order auto-correlation histogram of dot A under a 40 MHz pulsed excitation. Gray-colored region and red line indicate a background level and a fitted curve of the correlation histogram.

Cavities can modify the local density of optical modes, which enhances the spontaneous emission rate of dots [35]. From numerical calculation we obtain a cavity mode volume ($V$) of $0.66(\lambda/n)^3$. Combining this value with the measured $Q$ of 380 for mode M3, we determine a Purcell factor ($F_P = \frac{3}{4\pi^2}\left(\frac{\lambda}{n}\right)^3 \frac{Q}{V}$) of 44 at the maximum cavity intensity. To confirm the actual strength of the coupling between the quantum dots and the cavities, we performed time-resolved lifetime measurements. Figure 4(a) compares the lifetime of dot A to a second dot in the bulk. Dot A has a lifetime of 650 ps, while the bulk dot has a lifetime of 1.7 ns which lies within the uncertainty of the average bulk dot lifetime of 1.78±0.2 ns. The cavity-coupled dot shows a bi-exponential decay behavior ($\tau_{fast}$ = 650 ps and $\tau_{slow}$ = 1.8 ns). We attribute the slow decay component to occasional conversion to dark exciton states [40] and also to other uncoupled dots that potentially excite the cavity through non-resonant energy transfer [41, 42].

To verify that the reduced lifetime is a cavity effect, we performed lifetime measurements using 32 quantum dots from 20 different cavities. Figure 4(b) plots the measured lifetimes of all of these dots as a function of their detuning from the cavity mode M3. The distribution shows an enhanced spontaneous emission rate when the quantum dots are on resonance with mode M3, while the far detuned dots show a suppressed spontaneous emission rate, even lower than the averaged bulk dot rate, shown as a red dashed line. This behavior is consistent with a resonantly enhanced and suppressed Purcell effect [35]. We observe a lifetime as short as 400 ps for the fastest emitting quantum dot, corresponding to a Purcell enhancement of 4.4±0.5.

Together with spectral coupling, we consider spatial coupling, which multiplies a spatial mismatch term, $|\mathbf{E}(\mathbf{r})|^2/|\mathbf{E}_{max}|^2$, in the Purcell factor, where $\mathbf{E}(\mathbf{r})$ and $\mathbf{E}_{max}$ denote the electric field at the position of dots and the maximum field intensity [43]. The Purcell factor drops by a factor of 10 when the dot is only about 150 nm away from the field maximum (Supplement 4). In addition, the polarization mismatch between dots and cavities further reduces the Purcell factor [44]. These imperfect spatial and polarization matches cause the discrepancy between the calculated and measured Purcell effect, and explains why some dots have negligible Purcell effect even at near zero spectral detuning in Fig. 4(b).

Figure 4(c) shows the emission intensity of dot A as a function of pump power using continuous wave laser excitation, along with a bulk quantum dot for comparison. We choose the bulk dot that has a similar wavelength to dot A and higher brightness than average bulk dots (see Supplement 2). Well below saturation, dot A emits 80 times brighter than the bulk quantum dot. We attribute this intensity difference to the fact that a large fraction of the emitted photons from the bulk dot reflect back into the substrate due to total internal reflection. We note that in this measurement, dot A emits at a count rate that would ordinarily saturate our single photon detectors that have a minimum dead time of 4 µs, corresponding to a maximum count rate of 250 KHz. To avoid saturation we inserted a calibrated 20 dB attenuator in front of the detectors when measuring the saturation curve for dot A. The count rates in Fig. 4(c) are normalized by the transmission of the 20 dB attenuator. Because the bulk dot emits at a much lower intensity, we did not require an attenuator to obtain this saturation curve.

The solid lines in Fig. 4(c) represent numerical fits of the quantum dot intensities to a saturation curve given by

$$I = I_{max}\left(1 - e^{-\frac{P}{P_{sat}}}\right) \tag{1}$$

where $I_{max}$ is the maximum quantum dot emission intensity and $P$ and $P_{sat}$ are the pump power and the saturation power, respectively. From this fit we calculate a maximum emission intensity of 1.55±0.01 Mcounts per second for dot A, as compared to 19±0.2 Kcounts per second for the bulk dot. We expect higher $P_{sat}$ for the Purcell-enhanced dots, but dot A and the bulk dot do not show much difference in their $P_{sat}$. We attribute this to the different structures where these dots are embedded. More scattered and back reflected laser light would exist in the patterned membrane. However, we observe a difference in the $P_{sat}$ between the cavity-coupled dots with different Purcell factors (Supplement 5).

We measured the out-coupling efficiency by repeating the measurement of emission rate vs. pump power using a 50-ps pulsed laser with a repetition rate of 5 MHz. From this measurement we observed a maximum quantum dot intensity of 37±1 KHz, which corresponds to a photon detection probability of 0.74±0.02%. To estimate the collection efficiency through the first objective lens, we measured the transmission loss through each optical component. Our spectrometer has an efficiency of 42±2%, the coupling efficiency to the fiber is 48±4%, and we encounter additional losses due to lenses, mirrors, and beam splitters that further reduce the efficiency by 41±2%. These factors combine to give an overall efficiency of 8.3±0.9% for the collection system. We set the detector quantum efficiency to 20%, which further reduces the overall detection efficiency to 1.6±0.2%. Using these numbers we estimate that we collected 46±6% of quantum dot emission into the first objective lens (NA=0.7). If we consider non-zero $g^{(2)}(0)$ value, causing multi-photon events at high excitation power, we attain the multi-photon corrected efficiency of 36±5% [15] (see Supplement 3).

The dominant loss mechanism that limits the efficiency is due to the photons emitted in the down direction away from the objective lens. From numerical simulations we calculate that 62% of the light is emitted in the upward direction including back reflection from a bottom InP layer, and a fraction of 80% lies within the numerical aperture of the lens for mode M3. Also, the coupling efficiency for the cavity-coupled dot A is $\beta \approx 1 - \frac{\tau_c}{\tau_{uc}} = 0.77$, where $\tau_{uc}$ and $\tau_c$ are the lifetimes for the uncoupled (off resonant) and the coupled (on resonant) dots [35, 43]. These numbers give the collection efficiency of ~39%, which well explains the measured collection efficiency. This collection efficiency is higher than that from the previously reported optical horn structures that emit bright single photons at telecom wavelength [29] and therefore, could provide bright deterministic single photon sources for long-distant quantum key distribution [7, 45].

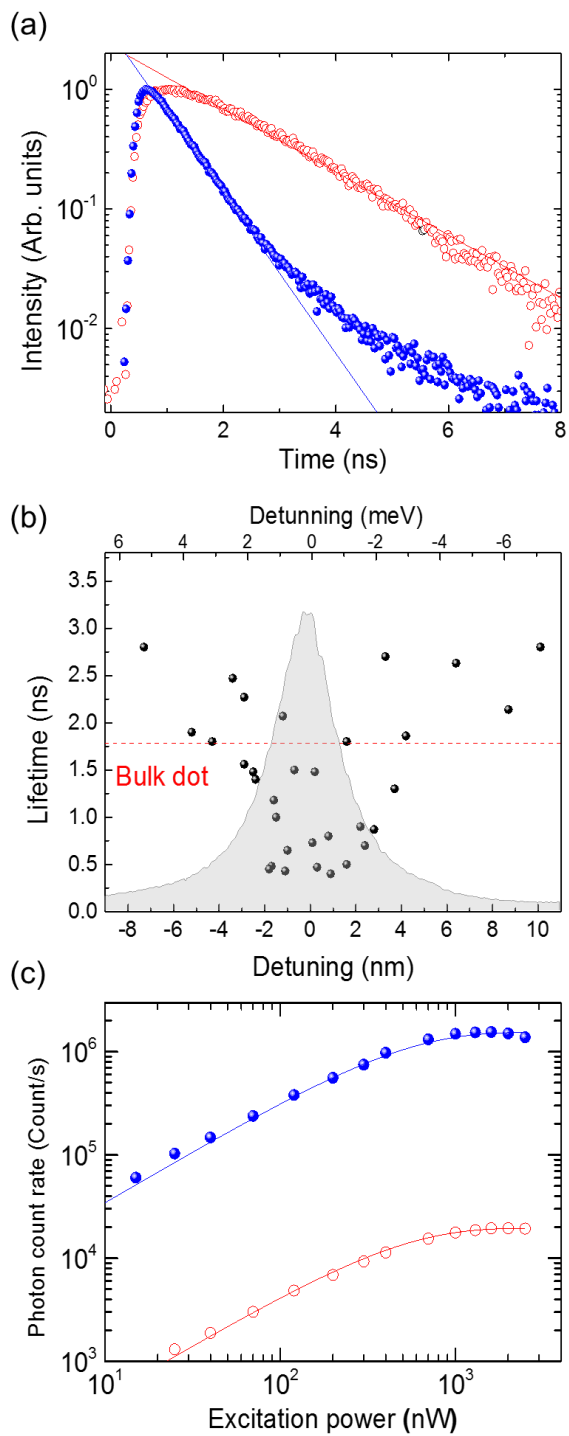

Fig. 4. (a) Decay curves of the cavity-coupled dot (blue-solid dots) and the bulk dot (red-empty dots). Measured data are fitted by single exponential decay functions (Solid lines). (b) Statistical distribution of lifetimes of individual cavity-coupled quantum dots. Red dotted line represents an average lifetime of bulk dots. Cavity mode M3 is shown in gray color. (c) Photon count rates of dot A (blue-solid dots) and the bulk dot (red-empty dots) as a function of excitation power. Solid lines are fitted curves for calculation of saturation intensity.

Many quantum information applications require indistinguishable photons that exhibit two-photon interference [1-3]. To investigate the indistinguishability of the source we performed a Hong-Ou-Mandel two-photon interference experiment (Fig. 1(b)). Figure 5(a) shows the two-photon interference result of dot A with parallel polarization. The correlation histogram consists of a series of 5 peaks, which we label 1-5. The 5 ns peak separation of these peaks corresponds to the path length differences between the two arms of the Mach-Zehnder interferometers (Fig. 1(b)). These 5 peaks are repeated every 25 ns due to the laser pulse repetition rate of 40 MHz. We used an average excitation power of 12 nW, corresponding to the saturation power of dot A for pulsed excitation. The center five peaks provide information about the indistinguishability of the source. For an ideal indistinguishable source, peak 3 which is centered around zero time delay should completely disappear due to two-photon interference, leading to an intensity ratio of 1:2:0:2:1. But for realistic sources a residual peak still exists due to imperfect temporal overlap, polarization mismatch of two-photon wavepackets, and also due to dephasing and timing/spectral jitter of the source [46, 47].

Figure 5(b) is a close-up of peak 3. For comparison, we also plot the two-photon interference results for orthogonally polarized two photons, attained by rotating the orientation of one of the polarization-maintaining fiber relative to the other. The graphs show a suppression of the coincidence events at zero delay time for the parallel polarization as compared to the orthogonally polarized photons. This suppression is the signature of two-photon quantum interference. At first we define the interference contrast given as $V = \frac{A_\perp - A_\parallel}{A_\perp}$, where $A_\perp$ and $A_\parallel$ are the total integrated areas under the two center peaks of the orthogonally and parallel polarized photons in Fig. 5(b). This results in the indistinguishable visibility $V = 0.18 \pm 0.01$. Dephasing and timing jitter cause this low visibility [47] and has a clear effect on the correlation curve which manifests itself as a sharp dip around zero delay for parallel-polarized photons.

With temporal post-selection method we calculate $g_{\parallel}^{(2)}(0) = 0.17 \pm 0.01$ by normalizing the parallel-polarized autocorrelation at zero delay by the averaged peak value of peaks 2 and 4. This value is below the limit of 0.5 for distinguishable single photons, while the orthogonally polarized photons show $g_{\perp}^{(2)}(0) = 0.49 \pm 0.02$, which is the expected value for distinguishable single photons. We calculate the post-selected interference contrast at zero delay as $\bar{V} = \frac{g_{\perp}^{(2)}(0) - g_{\parallel}^{(2)}(0)}{g_{\perp}^{(2)}(0)} = 0.67 \pm 0.02$. This number quantifies the degree of indistinguishability we can attain by post-selecting photons coincidences on timescales smaller than the coherence time ($\tau_2$).

The width of this dip gives the information of $\tau_2$. But in our experiments, the detector time response, the broad pulse width of the laser, and the length mismatch between the delay lines also influence on this dip, and therefore, the width of this dip is a convolution of these effects. We fit the curve for the orthogonal polarization to a simple form $f(\tau) = e^{-\frac{|\tau|}{\tau_1}}$, and fit the curve for the parallel polarization to a function of the form $f(\tau) = e^{-\frac{|\tau|}{\tau_1}}(1 - \upsilon e^{-\frac{2|\tau|}{\tau_{deph}}})$ to account for dephasing [48], where $\tau_1$ and $\tau_{deph}$ is the spontaneous emission rate and the dephasing time of the quantum dot, $\upsilon$ is the visibility of two-photon interference. $\tau_{deph}$ has a following relationship, $\frac{1}{\tau_{deph}} = \frac{1}{\tau_2} - \frac{1}{2\tau_1}$. We set $\tau_1$ equal to the value obtained from the lifetime measurement, while $\tau_2$ and $\upsilon$ are fitting parameters. We convolve the entire function with a Gaussian to account for the 200 ps time resolution of our avalanche photodiodes. From the fit we attain $\upsilon = 0.97 \pm 0.04$ and $\tau_2 = 150 \pm 29$ ps. We note that $\upsilon$ is within the margin of error for perfect visibility ($\upsilon = 1$), which suggests that the non-vanishing component of the measured data at zero time delay is primarily due to the finite time resolution of the detector. The blue-dashed line in Fig. 5(b) is the simulated curve for the indistinguishability we would obtain using ideal detectors with perfect time response.

Although the post-selection approach leads to higher indistinguishability, it expenses the brightness of single photons (Supplement 6). To overcome the low visibility without post-selection, and to produce high purity, on-demand indistinguishable single photons, we should reduce the dephasing rate and timing jitter [46]. Recently a number of works have demonstrated the highly indistinguishable single photon sources using quasi-resonant [14, 39, 49] or s-shell resonant excitations [16-18, 50] at near infrared wavelengths. Also applying electric bias can reduce the environment noise from the charges [18].

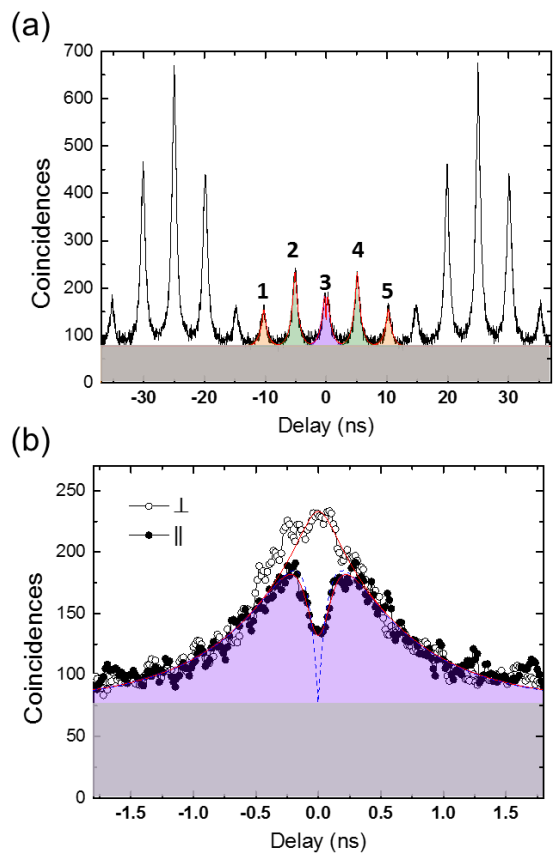

Fig. 5. Two photon interference measurement with Hong-Ou-Mandel setup for dot A. (a) Correlation histogram for parallel polarization. Five peaks labeled 1-5 every 25 ns represent the detection of two photons passing through different paths in two unbalanced Mach-Zehnder interferometers. Each peak is filled by different colors. (b) Close-up of the center peak for parallel (solid dots) and orthogonal polarizations (empty dots). Red lines are fitted curves, and the blue-dashed line is a simulated curve with an infinitely fast detector. Gray-colored region indicates a background level for the correlation histogram.

In addition to high brightness and indistinguishability, a well-defined polarization state of the quantum dot emission is also important in many quantum information process applications [3, 51, 52]. In particular, two-photon interference occurs when two photon wavepackets have the same polarization as well as wavelength, spatial, and temporal matching. Non-polarized singe photon emitters would require an additional linear polarizer to post-select one polarization state, resulting 50% loss of photons.

To characterize the polarization property of the cavity-coupled dots, we performed photoluminescence measurements. Fig. 6(a) shows the measured spectrum. In this sample we fortuitously found two bright quantum dots resonant with mode M3 denoted as dot 1 and dot 2, and two other bright dots resonant with mode M4 denoted as dot 3 and dot 4. Figure 6(b) shows the cavity emissions as a function of polarization at high pump power that saturates the quantum dots. The cavity modes show highly polarized emission. The polarization behavior agrees well with the expected polarization behavior shown in Fig. 2(b), where the polarization direction of modes M3 and M4 are orthogonal to modes M1 and M5. We do not clearly observe mode M2 in this figure due to its low intensity.

To confirm the polarization properties of quantum dots 1-4, we performed the same measurement at pump power well below the quantum dot saturation level. Figure 6(c) shows the results. The cavity-coupled dots 1-4 show highly polarized emission along the same direction as the cavity mode that they are coupled to. Figure 6(d) shows polar emission plots of modes M1 and M3, as well as dot 1 at 1312.5 nm as a function of polarization angle. We define a polarization ratio as $\rho=(I_{max}-I_{min})/(I_{max}+I_{min})$, where $I_{max}$ and $I_{min}$ are the maximum and minimum emission intensity. The modes M1 and M3, and dot 1 all show high polarization ratios of 0.93, 0.96, and 0.96 respectively. These results indicate that low-$Q$ mode M3 can produce linearly polarized single photons from the coupled dots. Therefore, this mode is very useful for linearly polarized, bright single photon sources.

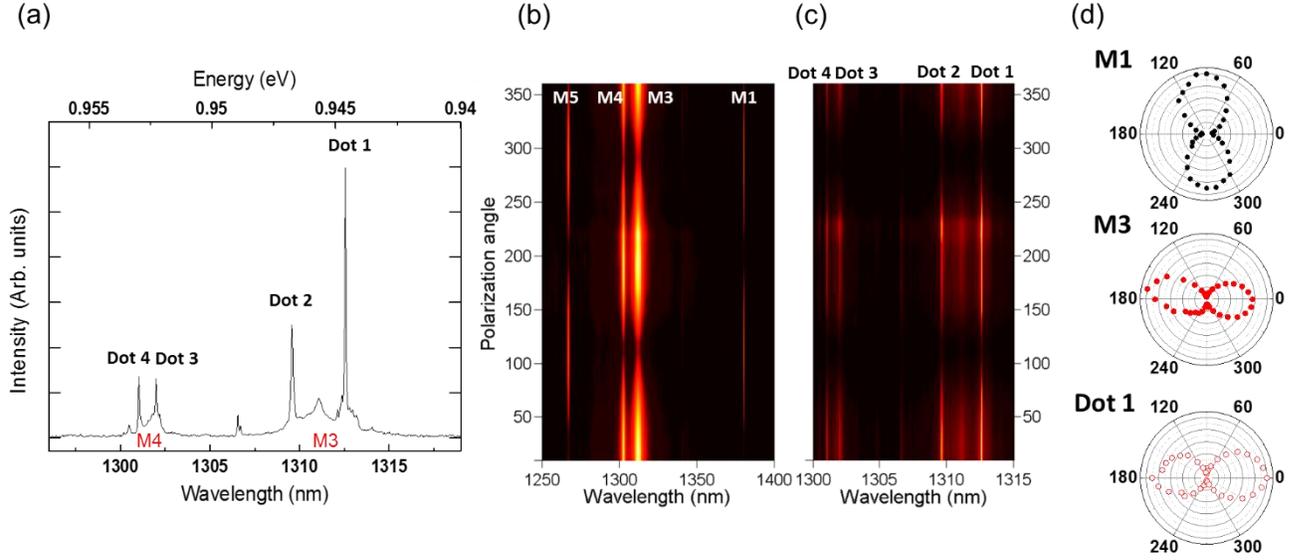

Fig. 6. Polarization measurement for cavity-coupled dots. (a) Photoluminescence spectrum of quantum dot emission coupled modes M3 and M4. Dots are denoted as dots 1-4. (b) Polarization angle scan of the cavity modes showing strong linear polarization. M3 and M4 modes show opposite polarization direction to M1 and M5 modes. (c) Polarization angle scan of the quantum dots coupled to M3 and M4 modes. Dots 1-4 show the same polarization dependence as that of M3 and M4. (d) Polar plots of the emission intensities of M1, M3, and dot 1 in (b) and (c) as a function of the polarization angle. They all have strong linear polarization ratios of 0.93, 0.96, and 0.96 for M1, M3, and dot1, respectively.

We have demonstrated two-photon interference from the cavity-coupled InAs/InP quantum dots that emit bright, highly polarized single photons at telecom wavelengths. These properties are essential for a broad range of applications in long-distance quantum communication. By utilizing low-$Q$ modes that are highly directional, we attained multi-photon corrected collection efficiencies as high as 36%, which is the highest value we are aware for telecom wavelength quantum dot single-photon sources. These low-$Q$ modes provide the additional advantage that they are relatively broad band and do not require highly precise matching of the quantum dot resonance to the cavity, and also induce Purcell effect in a broad range. The efficiency of our device could be further improved by introducing a distributed Bragg reflector mirror to re-direct the emission from the bottom of the cavity [53], which could potentially enable collection efficiencies as high as 80%. Furthermore, superconducting nanowire based single photon detectors with detection efficiencies over 90% are now available at telecom wavelength, which would enable even higher detection rates [54]. We could improve the indistinguishability using quasi-resonant [14, 39, 49] or resonant excitation schemes [16-18, 50], which have already been demonstrated to improve indistinguishability at near infrared wavelengths. We note that InAs/InP quantum dots could potentially emit single photons at 1.5 μm [27, 30], so our device and approach could be readily extended to this wavelength range for generating single photons at the *c*-band. Ultimately, our results show that InAs/InP quantum dots are promising candidates for producing indistinguishable photons for long distance quantum information applications [55, 56].

ACKNOWLEDGMENT

The authors would like to acknowledge support from the Laboratory for Telecommunication Sciences, the DARPA QUINESS program (grant number W31P4Q1410003), and the Physics Frontier Center at the Joint Quantum Institute.

# SUPPLEMENT INFORMANTION

1. Background levels in $g^{(2)}(\tau)$ measurement

For the correlation measurement, we used cooled InGaAs single photon detectors (ID230, Picoquant). Compared to conventional Si detectors, InGaAs detectors have a lower quantum efficiency of 20% and higher dark counts of ~200 Hz. We determined the background level in the correlation histograms in Fig. 3(b) and Fig. 5(a,b) as below.

The obtained correlation histogram ($g^{(2)}(\tau)$) is a product of two detector signals given by

$N_{m1} \cdot N_{m2} = (N_{s1} + N_{b1}) \cdot (N_{s2} + N_{b2}) = N_{s1} \cdot N_{s2} + N_{s1} \cdot N_{b2} + N_{b1} \cdot N_{s2} - N_{b1} \cdot N_{b2} \approx N_{s1} \cdot N_{s2} + 2N_{m1} \cdot N_{b2} - 3 N_{b1} \cdot N_{b2}$

$\approx N_{s1} \cdot N_{s2} + 2N_{m1} \cdot N_{b2}$

, where $N_{mi}$, $N_{si}$, and $N_{bi}$ indicate the measured signal, the quantum dot signal, and the background signal at detector i, respectively. Most background signal originates from the detector dark counts in our pulsed measurements.

Based on the fact that two detectors have a similar efficiency and $N_{m1} \gg N_{b2}$, the second term, $2(N_{m1} \cdot N_{b2})$, describes the background level in the measured histogram. To quantify the background level, we made detector 1 measure both the quantum dot and background signals while detector 2 only measure the detector dark counts by disconnecting the signal channel. We then calibrated the correlation data with time and took double this value.

2. Bulk quantum dot spectrum

In Fig. 4(c), we compared the intensity of the bulk dot and dot A. For this comparison, we chose one of bulk dots that has a similar wavelength to dot A and higher intensity than that of the average dots as shown in Fig. S1.

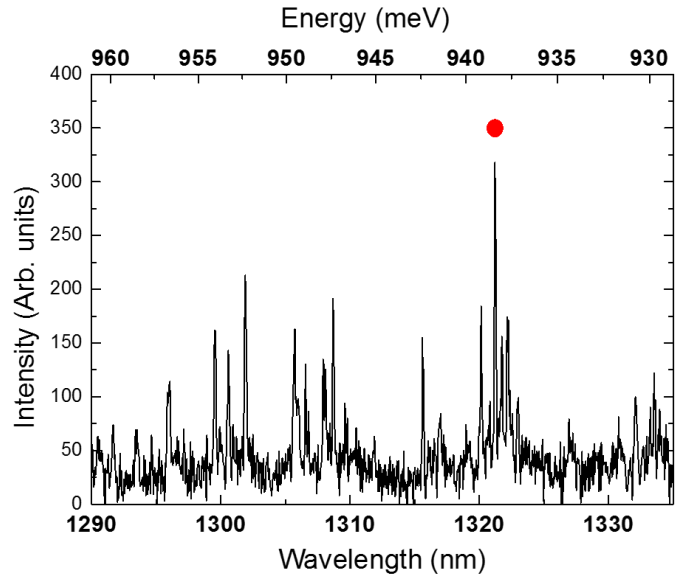

Fig. S1. Photoluminescence spectrum for bulk dots. The red dot indicates the bulk dot chosen for the comparison to the cavity-coupled dot A in Fig. 4(c).

3. Power dependence of $g^{(2)}(0)$

We investigated the excitation power dependence of $g^{(2)}(0)$ value of dot A. Figure S2(a) shows the measured correlation histograms at various excitation powers. The figure shows that the $g^{(2)}(0)$ values remain low at low power but start to increase near the saturation power ($P_{sat}$ =12 nW). The increased $g^{(2)}(0)$ values at high excitation power originates from the increased background emission by other dots and the enhanced cavity emission by nonradiatively coupled to other dots [1, 2]. To reduce the probability of multiple photon events, we should use low excitation powers, which could limit the brightness of single photon sources. We plot the $g^{(2)}(0)$ values and the measured collection efficiency as a function of power. Figure S2(b) shows the $g^{(2)}(0)$ value remains below 0.5 up to the maximum dot intensity. From non-zero $g^{(2)}(0)$ values, we correct the multi-photon events in the measured collection efficiency by multiplying the term, $(1-g^{(2)}(0))^{1/2}$ [3]. This reduces the collection efficiency from 46 % to 36 % at the maximum dot intensity.

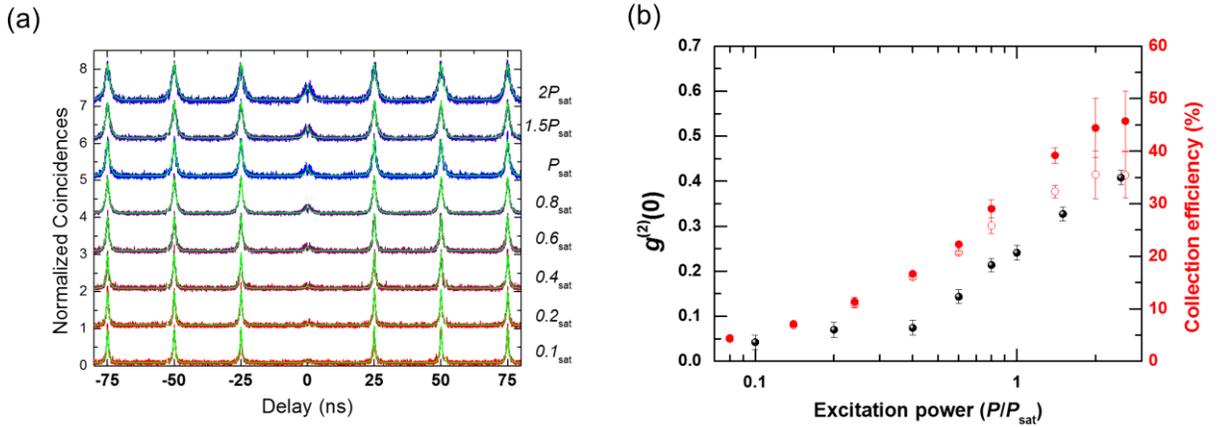

Fig. S2. (a) Photon correlation histogram at various excitation power levels. Green solid lines are fitted curves. (b) Plots of $g^{(2)}(0)$ values (black-solid dots), measured collection efficiency (red-solid dots), and multi-photon corrected collection efficiency (red-empty dots) as a function of excitation power.

4. Spatial coupling between cavities and dots

In our photonic crystal cavity with the measured $Q$ of 380 and the simulated mode volume ($V$) of $0.66(\lambda/n)^3$ for mode M3, we calculate the Purcell factor of 44 when we assume an ideal coupling between dots and cavities. However, the measured dots show one order of magnitude lower Purcell factors than the calculated value even at zero spectral detuning. This is mostly due to the spatial and polarization mismatches. To consider non-ideal spatial coupling, we multiply a spatial mismatch term, $|E(r)|^2/|E_{max}|^2$ in the Purcell factor, where $E(r)$ and $E_{max}$ denote the electric field at the position of dots and the maximum field intensity. Figure 3(a-c) shows the simulated spatial map of $|E(r)|^2/|E_{max}|^2$ for mode M3 and their intensity profiles along x and y directions. Mode M3 has a two maximum peaks at the middle of cavity, denoted as P1 and P2. Figure 3(b,c) shows that the Purcell factor drops by a factor of 10 when the dot is about 150 nm (85 nm) away from the field maximum along x (y) direction. This small size of localized modes makes the spatial coupling difficult, and becomes a main cause of the reduced Purcell factor in the experiments [4-7].

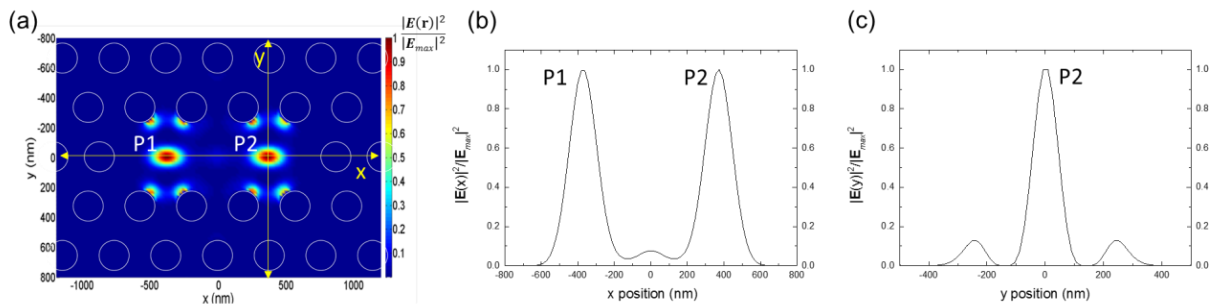

Fig. S3. (a) Simulated spatial map of $|E(r)|^2/|E_{max}|^2$ for mode M3. (b,c) Intensity profiles of $|E(r)|^2/|E_{max}|^2$ along (b) x and (c) y directions. Yellow lines indicate the positions for the curves in (b) and (c).

5. Purcell effect on cavity-coupled dots

Purcell effect enables dots to emit single photons with an enhanced spontaneous emission rate, and therefore, it also increases the maximum brightness of the dots by increasing the saturation power level. To investigate the influence of the Purcell effect on the saturation power of dots, we chose two cavity-coupled dot 1 and dot 2 in Fig. S4(a), which have different lifetimes of 650 ps and 1.8 ns, respectively. Figure S4(b) plots the integrated intensities of dot 1 and dot 2 as a function of excitation power. To record the dot intensity we used a charge coupled detector because the single photon detectors have a strong saturation effect due to their dead time and require additional calibration process, performed for Fig. 4(c) in the manuscript. From the fitting, we determine dot 1 and dot 2 have the saturation powers at 540±28 nW and 240±15 nW, respectively. Dot 1 shows 2.3 times faster lifetime and 2.8 times higher saturation power compared to dot 2. This enhanced spontaneous emission rate and saturation power by the Purcell effect would be helpful to reduce timing jitter and to generate bright single photons.

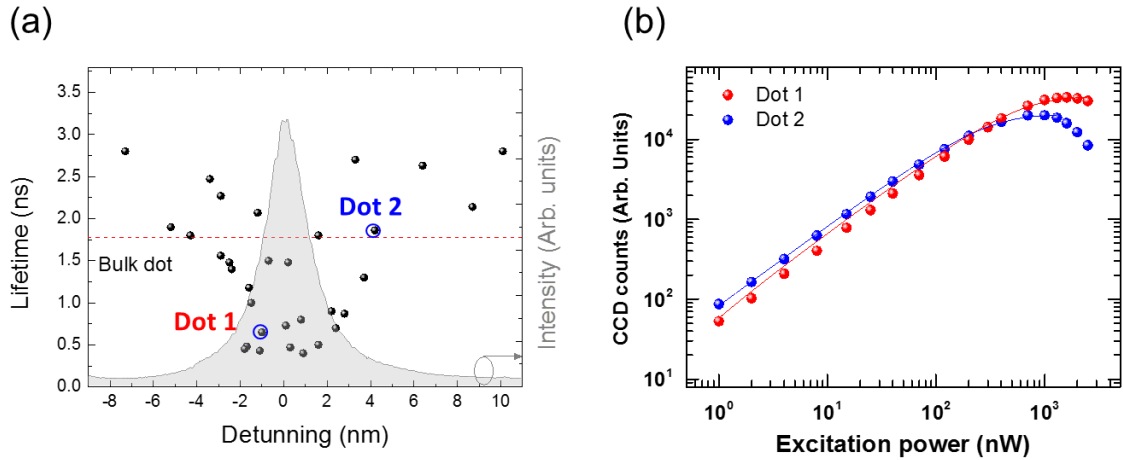

Fig. S4. (a) Statistical distribution of lifetimes of individual cavity-coupled dots. Dot 1 and dot2 indicate two dots with different lifetimes for comparison in (b). Cavity mode M3 is shown in gray color. (b) Plot of integrated intensities of dot 1 (red-solid dots) and dot 2 (blue-solid dots) as a function of excitation power. Solid lines are fitted curves for calculation of saturation intensity.

6. Indistinguishability vs. brightness

In Fig. 5(b), the cavity-coupled dot shows indistinguishable nature of single photons from two-photon interference. However, the strong dephasing limits the indistinguishable visibility of the dot and requires post-selection method to achieve high indistinguishability. Figure S5 shows that the indistinguishability increases as we reduce the time window for post-selection, while we lose the brightness of single photons. To achieve both high indistinguishability and brightness we can consider quasi-resonant [7-9] or s-shell resonant excitations [10-13] that reduce the influence of dephasing,

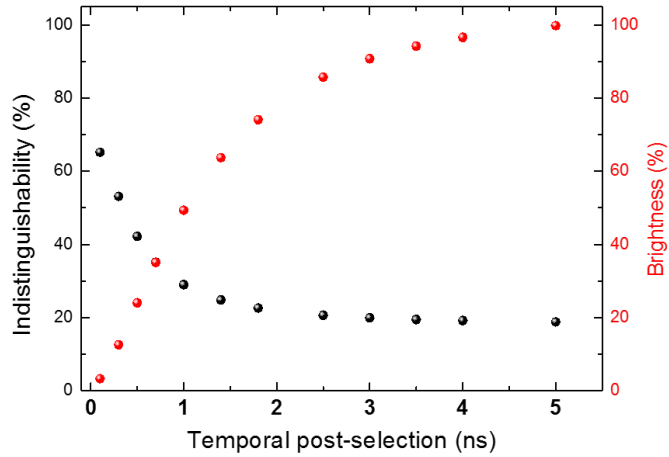

Fig. S5. Indistinguishability (black dots) and brightness (red dots) of single photons as a function of temporal post-selection in Fig. 5(b).